\documentclass{VRARWorkshop}

\usepackage[utf8]{inputenc}
\usepackage[hidelinks]{hyperref}

\usepackage{graphics}      
\usepackage{txfonts}
\usepackage{mathptmx}
\usepackage{color}
\usepackage{booktabs}
\usepackage{subcaption} 
\usepackage{textcomp}
\usepackage{wrapfig}
\usepackage{tikz}
\usetikzlibrary{positioning,shapes,shadows,arrows,automata,patterns,backgrounds,fit,calc}

\title{Novel Approach to Measure Motion-To-Photon and Mouth-To-Ear Latency in Distributed Virtual Reality Systems}

\authors{Armin Becher\VRARafftag{\ast}, Jens Angerer\VRARafftag{\dagger}, Thomas Grauschopf\VRARafftag{\star}}

\affiliations{
\begin{tabular}{ccc}
\VRARafftag{\ast} Technische Hochschule Ingolstadt & \VRARafftag{\dagger} AUDI AG & \VRARafftag{\star} Technische Hochschule Ingolstadt \\
Esplanade 10  & Auto-Union-Straße 1 & Esplanade 10 \\
85049 Ingolstadt & 85045 Ingolstadt & 85049 Ingolstadt\\
armin.becher@thi.de & jens.angerer@audi.de & thomas.grauschopf@thi.de
\end{tabular}
}

\abstract{%
Distributed Virtual Reality systems enable globally dispersed users to interact with each other in a shared virtual environment. In such systems, different types of latencies occur. For a good VR experience, they need to be controlled. The time delay between the user's head motion and the corresponding display output of the VR system might lead to adverse effects such as a reduced sense of presence or motion sickness. Additionally, high network latency among worldwide locations makes collaboration between users more difficult and leads to misunderstandings. To evaluate the performance and optimize dispersed VR solutions it is therefore important to measure those delays. In this work, a novel, easy to set up, and inexpensive method to measure local and remote system latency will be described. The measuring setup consists of a microcontroller, a microphone, a piezo buzzer, a photosensor, and a potentiometer. With these components, it is possible to measure motion-to-photon and mouth-to-ear latency of various VR systems. By using GPS-receivers for timecode-synchronization it is also possible to obtain the end-to-end delays between different worldwide locations. The described system was used to measure local and remote latencies of two HMD based distributed VR systems.
}


\begin{document}

\section{Introduction}
A simple prototype of a distributed Virtual Reality (VR) for remote collaboration in vehicle design was already built in 1997~\cite{Lehner.1997}. With new head-mounted displays (HMDs) like the \textit{HTC Vive} and \textit{Oculus CV1}, high-quality VR became affordable for many users worldwide. 


Although current distributed VR systems are significantly better than the prototype from 1997, there are still many problems to be solved before those systems can replace face-to-face contact. One of the challenges with Virtual Reality in general, and distributed VR in particular, are latencies. \textit{Normand et al.} stated, "[...] further study is required regarding the performance of the network communication, in particular regarding tolerance over the network's latency or the minimal ratio quality/latency required."~\cite{Normand.2012}. Other studies have shown that delays in VR systems have a big impact on user experience and need to be considered when designing distributed VR environments \cite{Brooks.1999,Meehan.2226March2003,St.Pierre.2015,Vaghi.}. Therefore, methods to measure and control latencies are required.



The time delay between movement and the corresponding display output is sometimes referred to as motion-to-photon latency~\cite{RajeshDesai.2014}. Locally this delay is perceived by a user if he or she makes a head movement and the corresponding display output is delayed by several milliseconds. If two or more users share the same virtual environment, remote latencies occur between different worldwide VR locations.   

Another type of delay in dispersed VR systems is audio latency. In this paper, audio latencies refer to the time delay between the generation of a sound signal until it is received at the local or remote output. Therefore, audio latency will further be referred to as mouth-to-ear latencies~\cite{Time.2003}. If users speak to each other inside a shared virtual environment, their voices need to be recorded and transferred over a network. With high mouth-to-ear latency, it is more difficult to communicate with each other naturally. For instance, if the audio delay between the communication partners is high, a user might start talking while another is already speaking.

The described measurement method allows scientists to further investigate the sources and effects of latencies in distributed VR systems. It can also help manufacturers and developers to evaluate VR hardware and applications.

In the next section of this paper, a quick overview on previous research work will be given. Afterwards, the novel measurement system will be presented. In the method description section, further technical details of the system will be provided. Then, the latency measurement results for two distributed VR systems will be shown. Afterwards, the accuracy of the described system gets examined. In the last section, a conclusion is given and plans for the future will be outlined.

\section{Previous work}

\subsection{Local motion-to-photon latency}
In 2014, \textit{S. Friston and A. Steed} characterized and compared three different approaches to measure motion-to-photon latency in VR systems~\cite{Friston.2014}. They named them \textit{Di Luca's Method}, \textit{Sine Fitting Method} and \textit{Automated Frame Counting Method}.

For the \textit{Sine Fitting Method} a video camera was used to measure local delays in a VR system. A 3D object was placed in the virtual scene and configured to follow the captured movements of a tracking system. A tracked object in the real world was attached to a pendulum and moved in a sinusoidal pattern. Both, the VR display showing the virtual object moving along the screen and the tracked object in the real world were captured with a video camera. Due to the harmonic oscillation of both objects a sine wave was fitted to the motion of each object. The resulting phase-shift between both waves was considered to be the motion-to-photon latency of the VR system~\cite{Steed.}.

In \textit{Di Luca's Method} instead of a video camera, two light sensing photodiodes were used to measure latency. One sensor was attached to the tracked object which was moved back and forth in front of a linear gradient from black to white. The other diode was attached to the VR display panel. A virtual gradient following the movement of the tracked object in the real world was displayed on the VR monitor. The captured luminance values of both photodiodes were recorded with the sound card of a PC. After the measurement, both resulting waveforms were cross-correlated to determine the motion-to-photon latency of the VR system~\cite{DiLuca.2010}.

To measure motion-to-photon delay when using \textit{Frame Counting}, a high-speed camera is required. Like the \textit{Sine-Fitting Method} a video camera is used to record the VR display and a tracked object in the real world simultaneously. A virtual 3D object following the movements of the tracked target is added to the VR scene. After capturing the high frame rate footage the time offset between the virtual and the tracked object can be determined. For example, if the real object reaches the peak position of the pendulum swing, the video frames until the virtual object gets to the corresponding position in the virtual world can be counted. 
Because manually counting frames is time-consuming and error-prone, an automatic \textit{Frame Counting} application has been developed by \textit{S. Friston and A. Steed}~\cite{Friston.2014}. Instead of manually analyzing the video footage, \textit{MATLAB} scripts were used to compute end-to-end latency automatically. 


Another measurement setup described by \textit{Raaen et al.} was designed to measure the motion-to-photon delay of HMDs and smartphones~\cite{Raaen.2015}. They used two light sensors, a laser pen, and an oscilloscope to detect the delay of a VR system. One diode was attached to the VR display. The laser pen was mounted on the HMD or smartphone and pointing towards another diode next to the VR device. Both diodes were connected to an oscilloscope. If the VR system was rotated in one direction, the VR display switched from white to black, and the laser pen did not hit the photosensor anymore. The latency was determined by measuring the time difference until both diodes switched from bright to dark.

\subsection{Remote motion-to-photon latency}
To measure the delay between different dispersed virtual environments, timecode synchronization is needed. \textit{Purdy et al.} described a hardware system to measure latencies in distributed flight simulators~\cite{Purdy.1998}. With the described system, it was possible to determine when a pilot in one simulator perceived virtual aircraft movements of another simulator. 

In 2009, \textit{Roberts et al.} introduced a method to measure the end-to-end latency of distributed collaborative environments by using two video cameras~\cite{Roberts.2009}. With an initial frame lock (genlock) the timecodes of both cameras were synchronized before actually taking a measurement, after which \textit{Manual Frame Counting} was used to determine the end-to-end latency between different locations. 

\subsection{Mouth-to-ear latency}
In most distributed virtual environments users communicate with each other using microphones and output devices such as headphones or speakers. If noise canceling headphones are used, it is necessary to feedback the user's voice because otherwise he or she would not hear his or her voice while speaking. In that case, measuring the local mouth-to-ear latency is also important because hearing the own voice with a high delay can be distracting.

\textit{Agastya et al.} investigated the mouth-to-ear latency in  VoIP clients and described a simple method to measure those delays~\cite{Agastya.2009}. First, they generated an input signal. This signal was split into two identical signals which were fed into a laptop and the caller machine's mic-in jack. The output of the callee machine's headphone jack was also fed into the laptop. Later a signal was generated, and the offset between generation and headphone output was compared to calculate the mouth-to-ear latency.

\section{Overview}

\begin{wrapfigure}[29]{r}[0pt]{0.46\columnwidth}
	\centering
	\includegraphics[width=0.46\columnwidth,trim={4cm 0 0 0},clip]{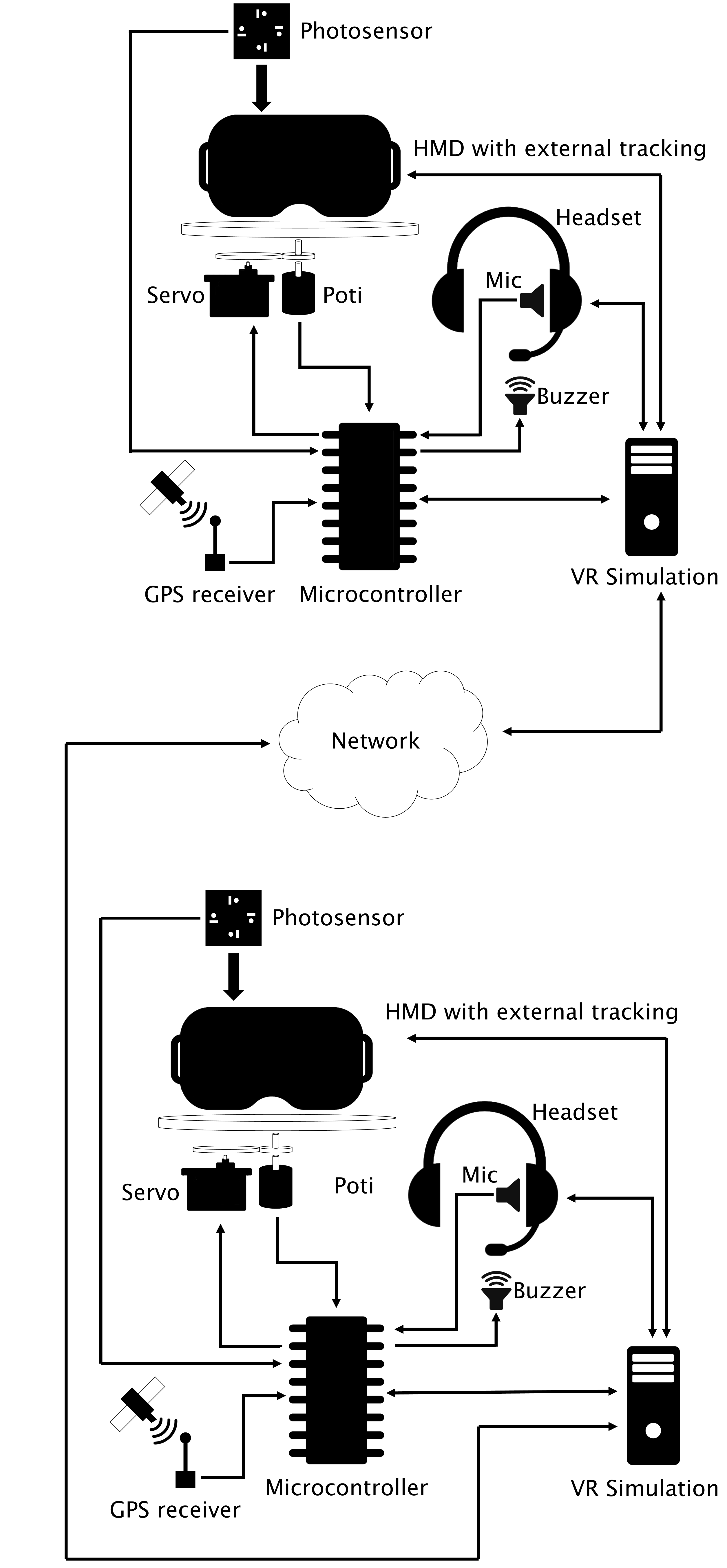}
	\caption {Latency Measurement setup}
	\label{fig:MeasurmentSetup} 
\end{wrapfigure}

In this paper, an open-source system to measure local and remote motion-to-photon latency will be presented. Compared to existing measurement techniques, the described method is specially designed for distributed VR systems. It will help to efficiently evaluate different setups regarding motion-to-photon and mouth-to-ear latencies.

The game-engine \textit{Unity 3D} is used to simulate the virtual world. The source code of the project can be downloaded from \textit{GitHub}
\footnote{\url{https://github.com/vr-thi/VRLate}}. 

\subsection{Motion-to-photon latency}
Motion and the corresponding display output of a VR system are captured by a potentiometer and a photosensor which are both connected to a microcontroller. An object tracked by the VR system is mounted on the potentiometer and automatically rotated back and forth. The motion is controlled by a servo motor connected to the microcontroller. By monitoring the horizontal rotation with the potentiometer, it is possible to determine the rotary position of the real object at every interval.

To get the delay between the rotation of the real object and the corresponding display output, the latest angle captured by the motion-tracking system is presented on the VR display. The displayed value is converted into a brightness code so that it can be read with the microcontroller. By attaching the measurement unit's photosensor to the display, it is possible to capture this brightness code and therefore the angle which was used by the VR system to render the current frame. 
The perceived motion-to-photon latency is determined by comparing both signals with each other using cross-correlation.

\subsection{Mouth-to-ear latency}
To determine the mouth-to-ear latency with the described measurement setup, a low-cost piezo buzzer and a microphone are used. Right before starting the measurement the microcontroller activates the buzzer for a short period. At every interval, the microcontroller checks if the attached microphone detects any sound. Then, the mouth-to-ear latency can be determined by counting the intervals between the beginning of the measurement and the received sound-impulse.

\subsection{Remote latency}
If the delay between multiple VR systems needs to be measured, synchronization is required. In the presented method, GPS receivers are used to synchronize multiple microcontrollers and measure delays in distributed VR systems. Figure \ref{fig:MeasurmentSetup} shows how the described technique can be used to measure local latency and delays between two remote VR locations. 

\subsection{Comparison to previous work}
Compared to camera-based methods~\cite{Steed.,Friston.2014,Roberts.2009}, the use of photodiodes and a potentiometer makes it easy to evaluate latencies of HMD systems. For all measurement methods using a video camera, it is necessary to remove the lenses from the HMD to capture the output of the VR display. If the motion-to-photon latency of the HMD itself should be measured it is required to attach the whole HMD to a pendulum. To capture the display, the camera needs to be attached to the pendulum as well. With the measurement system described in this work, there is no need to remove the lenses, and the end-to-end delay of a system can be determined quickly. This will help hardware and software developers to continuously evaluate changes to the VR system concerning end-to-end latencies.

With the proposed GPS clock synchronization method, it is possible to accurately measure remote latencies in distributed VR systems using cheap and available hardware components. This approach has advantages compared to the camera synchronization mechanism described by \textit{Roberts et al.}~\cite{Roberts.2009}. First of all, it is not necessary to synchronize both measurement devices by connecting them to a shared location. Another benefit of the GPS synchronization technique is that it does not suffer from noticeable time offsets. Although \textit{Roberts et al.} did not report any drifts, it is not clear for how long two cameras can run synchronized without recalibration. Regarding synchronization accuracy and granularity, the proposed method also surpasses camera-based methods. While the timecode of external microcontrollers can be precisely synchronized to differ less than 1~ms from each other, \textit{Roberts et al.} reported that the camera synchronization method has a synchronization granularity of one frame which leads to measurement accuracy of approximately 20~ms~\cite{Roberts.2009}.

The described method combines the use of photodiodes and a potentiometer to determine the motion-to-photon latencies in VR systems and is technically comparable to the method described by \textit{Di Luca}~\cite{DiLuca.2010}. Additionally, the mouth-to-ear latency is captured with a similar setup as described by \textit{Agastya et al.}~\cite{Agastya.2009}. The use of external measurement hardware and GPS synchronization makes it possible to quickly and accurately measure remote end-to-end latencies in distributed VR systems. The hardware and software tool is provided as an open-source project and allows others to use and enhance it to their specific needs.

\section{Method Description}

\subsection{Rotation platform}
For the measurement setup, a potentiometer is used to continuously trace the rotation angle of the tracked object (e.g., an HMD). A servo connected to the microcontroller controls the platform rotary motion. With this setup, steady back-and-forth movements of the tracked object are possible.

\subsection{Photosensor}
\label{sec:photosensor}
The photosensor is used to monitor display outputs and captures the rotary angle of the tracking system encoded in a brightness code.

\begin{figure}
	\centering
	\begin{subfigure}{.45\textwidth}
		\centering
		\includegraphics[height=5cm]{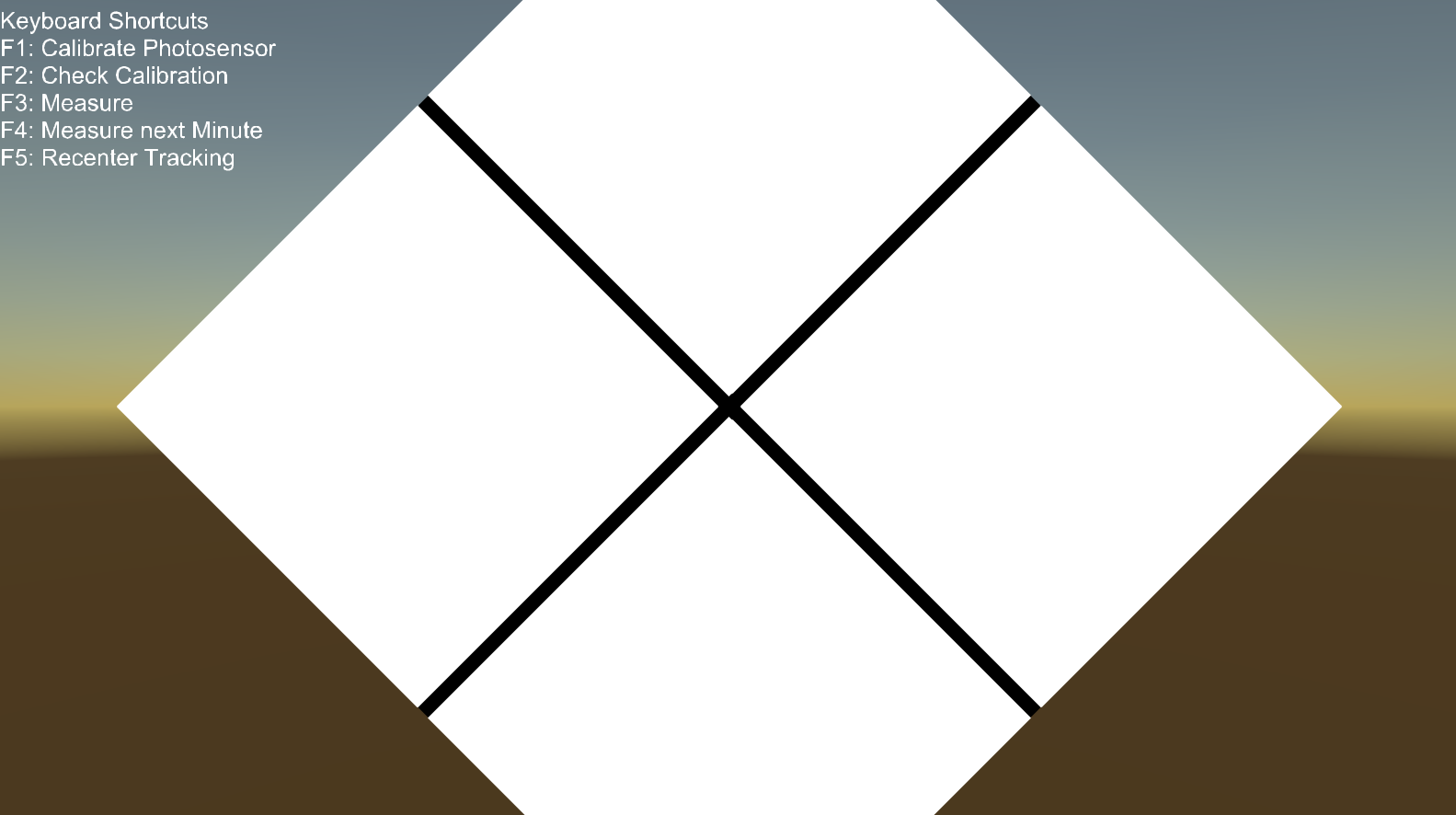}
		\caption {Brightness code}
		\label{fig:BrightnessCode} 
	\end{subfigure}
	\begin{subfigure}{.45\textwidth}
		\centering
		\includegraphics[height=5cm]{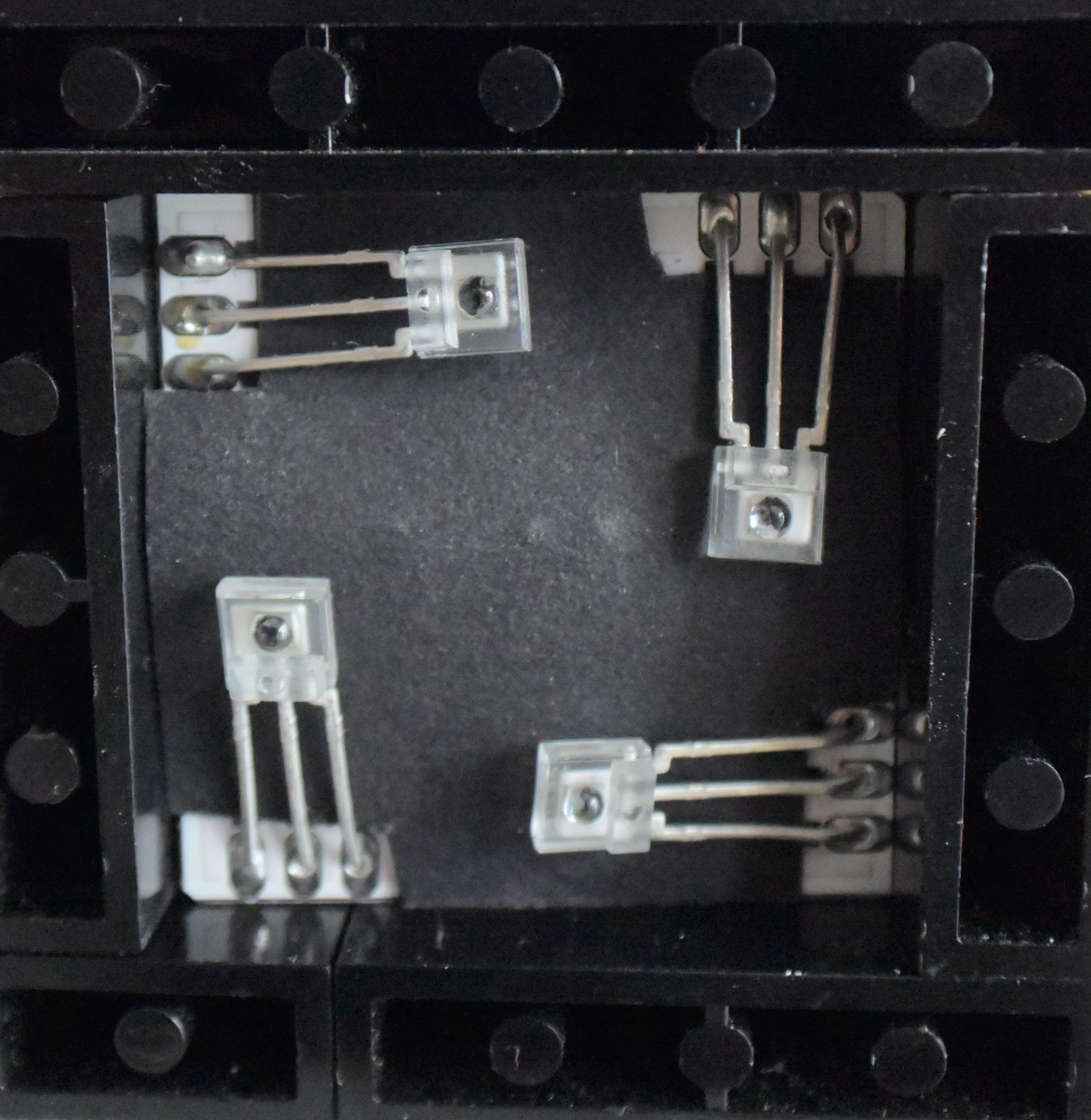}
		\caption {Photosensor}
		\label{fig:Photosensor} 
	\end{subfigure}
	\caption {Measure display output}
	\label{fig:diplayoutput} 
\end{figure}

During measurement, right before the rendering process, the VR system outputs the last captured horizontal rotation angle translated into a brightness code. To get a resolution of 4096  different values, four areas on the display are used to output a number in the octal numeral system. By capturing each area with a light-sensitive sensor connected to the microcontroller, it is possible to get the rotation angle output of the VR system at every interval. Figure \ref{fig:BrightnessCode} shows the displayed brightness code of the VR system. 
The octal numeral system is used because the analog signal of the photosensors is noisy and a bit of headroom between the different brightness levels is needed.

The whole photosensor consists of four \textit{TSL250R} photodiodes which are very similar to those used in \textit{Di Luca's Method}~\cite{DiLuca.2010}. According to the datasheet, the sensors have an output rise-time of about 260 $\mu s$~\cite{TAOSInc..2001}. Figure \ref{fig:Photosensor} shows the used photosensor. With the HMDs used in this work, it is not possible to directly attach the photosensor to the VR display without the need to disassemble the hardware. HMDs use optical lenses in front of the display system to expand the field of view. Without any light path correction, it is not possible to get a sharp image of the displayed frame on a flat surface.

To read the brightness code of the VR system on HMDs, an additional optical lens is necessary. For the measurement setup, convex lenses built into the low-cost VR glasses \textit{Google Cardboard} where directly attached to the lenses of the HMD to correct the light path of the VR display. Attaching the photosensor a few millimeters away from the correction lens makes it possible to read the displayed brightness code.

\subsection{Audio Latency Detector}
To detect the mouth-to-ear latency of VR systems, a combination of a piezo buzzer and a microphone is used. 

By attaching the buzzer to an analog output pin with pulse-width modulation (PWM) capability of the microcontroller, a square wave signal is generated to get a pure tone. For the described setup an audio frequency of 4~kHz is used since at around this frequency the sound pressure output of the used piezo buzzer is at its maximum. The buzzer is attached to the microphone of the VR system.

To detect the generated tone at the speakers of the VR system, a cheap sound detection model is used. The \textit{SparkFun Sound Detector}\footnote{\url{https://github.com/sparkfun/Sound_Detector}} consists of a simple electret microphone, a voltage feedback amplifier and a voltage comparator. When the sound pressure at the detection module reaches a specific threshold, the digital output pin of the module is set to high. One digital pin of the microcontroller is used as input pin and waits for a rising signal of the sound detection module. When a high signal is detected, the elapsed intervals since activating the buzzer are summed up and represent the measured mouth-to-ear delay.

\subsection{Microcontroller} 
\textit{Teensy~3.2} microcontrollers are used as central units of the described measuring system. With the built-in 16~Mhz crystal 
continuous measurements with a stable frequency are possible. Since the crystal clock accuracy is unspecified, a worst-case quality for quartz crystals of $\pm$100~ppm is assumed.
With the assumed worst-case accuracy, it is possible to take synchronized measurements with a maximum time deviation of $\pm$0.5~ms over a timespan of 5~s.

The photosensors and the potentiometer are connected to the analog inputs of the microcontroller. The built-in analog-to-digital converter (ADC) of the \textit{Teensy~3.2} is used to capture all sensor inputs and store them in the microcontroller's Random Access Memory (RAM). All analog-to-digital conversions of the connected sensors take less than 1~ms. 

With the described system, measurements are taken with a frequency of 1000~Hz. This frequency was chosen because of the built-in OLED displays of current HMDs. Those displays show a new frame for about 1-2~ms until the display turns black again. With intervals of 1~ms, it is possible to measure the display output at every frame.

Theoretically, even higher frequencies are possible, but for the described VR scenarios 1~ms intervals proved to be fast enough. Capturing with higher frequencies would also require additional RAM space.

\subsection{Synchronization}
To measure the delay between different locations, synchronized timecodes are needed. 

For the described setup two \textit{u-blox NEO-6M} modules continuously transmit GPS data to the connected microcontrollers. This data contain a \textit{UTC timestamp} which is used for synchronization. Besides the serial connection for the GPS data, the \textit{u-blox NEO-6M} has another pin which outputs a very accurate time pulse signal that rises every full second with a root-mean-square accuracy of 30~$\mu s$~\cite{UBloxAG.2011}. A UTC timestamp can be transmitted to the controllers for a synchronized start of measurement. All globally-distributed systems will wait until then and start taking measurements simultaneously. Compared to other timecode synchronization techniques, GPS receivers are a lot more precise. One of the drawbacks though is that the GPS antenna needs to be placed outdoors or close to a window to work. For the proposed system a long wire was used to connect the measurement system with the signal of the GPS module.



\subsection{Data evaluation}
After measurement, the raw data captured by the microcontrollers are transmitted to the VR simulation computer. Using \textit{Unity 3D} scripts, the photosensor data of all four photosensors are decoded into the corresponding decimal number representing the rotation output of the VR system.
The recorded potentiometer and photosensor data are cross-correlated with each other. 
The lag with the maximum correlation coefficient represents the motion-to-photon latency of the VR system.
\begin{figure} [h!]
	\centering
	\includegraphics[width=1\columnwidth]{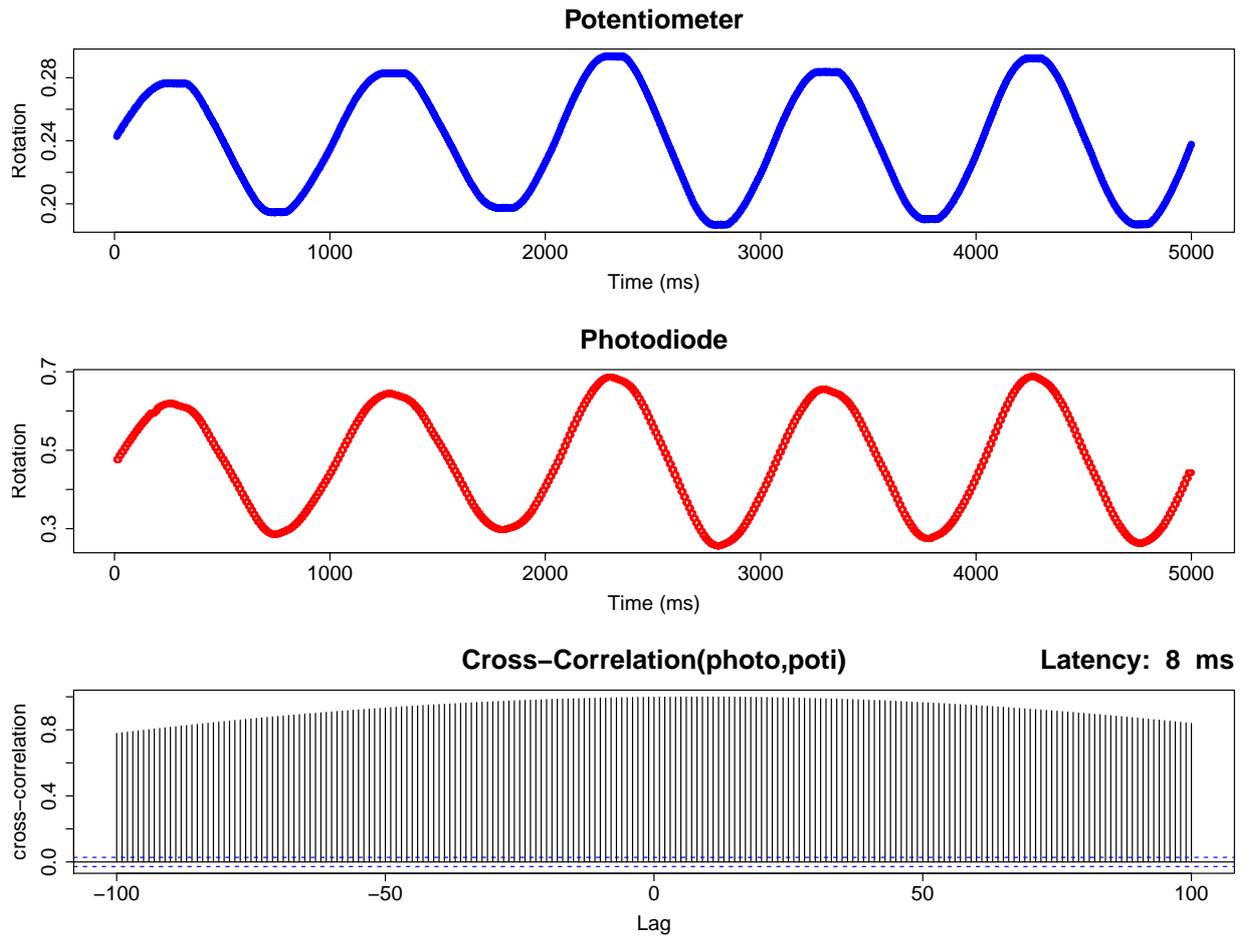}
	\caption {Motion-to-photon latency of the Oculus Rift CV 1}
	\label{fig:DataEvaluation} 
\end{figure}

Figure \ref{fig:DataEvaluation} shows an example of the measured motion-to-photon latency of the \textit{Oculus Rift CV1}. The blue line is the captured rotation of the potentiometer and the red line the corresponding display output. For example, in figure \ref{fig:DataEvaluation} the maximum correlation coefficient and therefore the estimated motion-to-photon latency is at 8~ms.

The microcontroller computes the mouth-to-ear latency with the attached microphone. Counting the intervals (1000~Hz) between the start of measurement and sound detection is the local or remote mouth-to-ear latency of the VR system in milliseconds.

\section{Measure Latency}
In the following sections, the measured motion-to-photon and mouth-to-ear latency of a simple distributed VR system will be described.

\subsection{Setup}
Two HMDs were used for latency measurements: the \textit{HTC Vive} and the \textit{Oculus Rift CV1}. Both VR devices were used with the respective external tracking system and got attached to computers with slightly different hardware specifications. To better distinguish between both configurations they will further be referred to as \textit{OCR} (PC with \textit{Oculus Rift CV 1}) and \textit{VIVE} system (PC with \textit{HTC Vive}).
The hardware specifications of both systems are listed in table \ref{tb:vivePC} and \ref{tb:ocrPC}. 

\begin{table}[h!]
	\begin{minipage}{0.5\textwidth}
		\centering
		\caption{\textit{VIVE} PC specifications}
		\label{tb:vivePC}
		\begin{tabular}{@{}ll@{}}
			\hline
			CPU & Intel Core i7-6700\\
			RAM    & 16 GB \\
			OS  & Windows 10 - 64 bit \\
			Graphics Card & GeForce GTX 970 \\
			\hline
		\end{tabular}
	\end{minipage}
	\hfill
	\begin{minipage}{0.5\textwidth}
		\centering
		\caption{\textit{OCR} PC specifications}
		\label{tb:ocrPC}
		\begin{tabular}{@{}ll@{}}
			\hline
			CPU & Intel Xeon E5-1620 \\
			RAM    & 8 GB \\
			OS  & Windows 10 - 64 bit \\
			Graphics Card & GeForce GTX 970 \\
			\hline
		\end{tabular}
	\end{minipage}
\end{table}

The VR simulation was running in \textit{Unity 3D} version 5.5.2f1. The displayed scene was empty except for the default \textit{Unity Skybox} and a tracked virtual camera. V-Sync was activated on both VR systems. 

For audio communication between both VR stations the audio chat tool \textit{Mumble} version 1.2.19 was used. \textit{Mumble} requires the server process \textit{Murmur}. The server was running on the \textit{OCR} PC (see table \ref{tb:ocrPC}). \textit{Philips SHP1900} headphones got plugged into the audio jack of the \textit{HTC Vive}. As audio input device the microphone of the \textit{Vive} was used. For the \textit{Oculus Rift} system the built-in microphone and headphones of the \textit{HMD} were used.

Both VR systems were directly connected with each other via a CAT6 network cable. With the built-in network card, a maximum transfer rate of up to 1~Gbit/s is possible. 
The round-trip time between both stations was less than 1~ms (average result of 10 ICMP echo requests). To synchronize the rotation of both VR systems the \textit{Unity High-Level API} (\textit{HLAPI}) was used. In the \textit{Unity Network Transform} component the maximal possible send rate of 29 updates per second was chosen\footnote{\url{https://answers.unity.com/questions/1431826/network-transform-send-rate-send-interval.html}}. 

\subsection{Results and Discussion}
Measurements were taken 10 times on each setup. Rotation data was recorded over a timespan of 5~s. The results are summarized in table \ref{tb:ResultsMotionToPhoton} and \ref{tb:ResultsMouthToEar}.

\begin{table}[h!]
	\begin{minipage}{0.5\textwidth}
		\centering
		\caption{Measured motion-to-photon latencies}
		\label{tb:ResultsMotionToPhoton}
		\begin{tabular}{@{}lllll@{}}
			& Avg     & Min      & Max & SD \\
			\hline
			\hline
			\textit{OCR}                & 5.8~ms&     3~ms &9~ms    & 0.5    \\
			\textit{VIVE}            & 5.1~ms&    1~ms &10~ms    & 2.7    \\
			\textit{O} $\to$ \textit{V}    & 49~ms    &    42~ms&55~ms    & 3.8    \\
			\textit{V}  $\to$ \textit{O} & 27.8~ms&24~ms     &33~ms    & 2.4    \\
			\hline
		\end{tabular}
	\end{minipage}
	\hfill
	\begin{minipage}{0.5\textwidth}
		\centering
		\caption{Measured mouth-to-ear latencies}
		\label{tb:ResultsMouthToEar}
		\begin{tabular}{@{}lllll@{}}
			& Avg     & Min      & Max    & SD  \\
			\hline
			\hline
			\textit{OCR}                &100~ms    &93~ms     &106~ms   &    3.7 \\
			\textit{VIVE}            &112.4~ms &112~ms &113~ms & 0.5 \\
			\textit{O} $\to$ \textit{V}    &144.1~ms&132~ms&151~ms   & 5.6 \\
			\textit{V}  $\to$ \textit{O}    &168.3~ms&157~ms&175~ms   & 5.4 \\
			\hline
		\end{tabular}
	\end{minipage}
\end{table}


The measured local motion-to-photon latency of both VR systems was very low. There was no significant difference between both systems. Both results are similar to the motion-to-photon latency of the \textit{Oculus Rift DK2} measured by \textit{Raaen and Kjellmo} in 2015~\cite{Raaen.2015}. However, this is only true if V-Sync was deactivated for the \textit{Oculus Rift DK2}. When V-Sync was active, \textit{Raaen and Kjellmo} measured an average latency of 41~ms for the \textit{Oculus Rift DK2}~\cite{Raaen.2015}. The results of this paper indicate that the newer generation of HMDs reaches a low motion-to-photon latency even if V-Sync is activated.

The measurement results in figure \ref{fig:DataEvaluation} suggest that extrapolation is used to achieve such low latencies. Comparing the turning points of the potentiometer and those of the display output in figure \ref{fig:DataEvaluation} shows that the recorded platform motion is flatter at the extremes than the movements recorded at the display. This indicates that the tracked rotation gets extrapolated and leads to overshooting.

The remote latency between both VR systems was high considering the fact that both stations were only a few meters apart and directly connected to each other. One reason for that might be that the network send rate of both systems was set to 29 updates per second. 

The latency difference between both directions was about 20~ms. This was unexpected since the measured local latency for both systems was more or less the same (see table \ref{tb:ResultsMotionToPhoton}). Maybe the difference is caused by differences in the used VR SDKs (\textit{Oculus} for the \textit{Oculus Rift CV1} and \textit{OpenVR} for the \textit{HTC Vive}). The tracking data of the \textit{HTC Vive} might have been extrapolated before they were sent over the network and therefore the measured delays from the \textit{Vive} to the \textit{OCR} is lower. However, this is only a vague assumption that requires further investigations.

To record the local mouth-to-ear latency of a VR system, the microphone was directly rooted to the loudspeaker using the standard \textit{Windows} volume mixer. The local measured audio delay differed only slightly between the measurements. 

The latency results for the \textit{HTC Vive} system are about 10~\% higher than those of the \textit{Oculus Rift}. Interestingly, the local mouth-to-ear delays are quite stable for the \textit{HTC Vive} system, whereas the latency results for the \textit{OCR} differed from each other up to 13~ms. 





\section{Evaluation}
In the following sections, the accuracy of the described method will be evaluated. Like in the previous section, all measurements were performed ten times.

\subsection{Motion-to-photon latency}
To determine whether the described method measures the expected motion-to-photon latency of a VR system, a frame delay mechanism was implemented in Unity. The brightness codes are not immediately rendered, but stored in a queue. The length of the queue determines how many frames the display output should be delayed. Before the rendering process, the current brightness code is added to the queue, and the oldest value in the queue gets displayed. The \textit{HTC Vive} was used to evaluate the accuracy of the motion-to-photon latency measurement setup. The \textit{HTC Vive} uses a 90~Hz display, and therefore a delay of one frame should increase the end-to-end delay by approximately 11.11~ms.

\begin{table}[h!]
	\centering
	\caption{Delayed motion-to-photon latencies}
	\label{tb:Measurement Motion-To-Photon latency}
	\begin{tabular}{@{}lllll@{}}
		& Avg     & Min      & Max & SD \\
		\hline
		\hline
		Baseline        & 5.1~ms    &     3~ms     &9~ms  & 2.6~ms \\
		One frame delay    & 18.1~ms    &    10~ms     &25~ms & 5.2~ms    \\
		Five frames delay    & 63.7~ms    &    56~ms    &68~ms & 4.1~ms    \\
		Ten frames delay    & 120.3~ms    &    114~ms    &124~ms & 2.9~ms    \\
		\hline
	\end{tabular}
\end{table}

Table \ref{tb:Measurement Motion-To-Photon latency} summarizes the results. The results show that the measurement system can detect the artificially added latency. Based on a system delay of 5.1~ms, the expected delay for one, five and ten frames is 16.2~ms, 60.7~ms, and 116.2~ms. The measured values in table \ref{tb:Measurement Motion-To-Photon latency} 
differ by a maximum of 4.1~ms from the expected system delays.

\subsection{Mouth-to-ear latency}
To evaluate the accuracy of the described mouth-to-ear latency measurement method, the used buzzer was directly attached to the microphone of the measurement system. With this configuration, the audio delay is expected to be close to zero milliseconds. The time delay between the generation and detection of the sound signal was measured with the microcontroller. Between the tone output and the measured signal passed $401.6\pm0.5\mu s$. An average delay of less than half a millisecond and the very low standard deviation of $0.5\mu s$ show that the mouth-to-ear delay can be measured very precisely with the described system.

\subsection{Clock synchronization}
With the following method, the GPS synchronization mechanism described in this work was evaluated. Both measurement systems were placed right next to each other, and two digital ports of both microcontrollers were connected with each other via a short cable. After the devices were synchronized to GPS time, both got a command to start measuring in the next minute. One microcontroller logged the start of measurement in microseconds and waited for a hardware interrupt on the input pin connected to the other measurement system. As soon as the other microcontroller started to measure, the input pin of the first microcontroller was set to high. Within the hardware interrupt routine of the first microcontroller, another timestamp was acquired. The difference between both timestamps was as low as $15.2\pm1.7 \mu s$. This shows that GPS clock synchronization works as expected.

\section{Conclusion and future work}
The presented measurement setup makes it easy to evaluate the motion-to-photon and mouth-to-ear latency of different distributed VR systems. 

Measuring latency of current VR hardware showed some interesting results. 
It seems that a very low local \textit{motion-to-photon} latency can be achievable with current HMDs even if V-Sync is activated. For the \textit{Oculus Rift DK2} \textit{Raaen et al.} concluded that they "[...] found no way of satisfying both the delay requirement and the requirement to use vertical synchronization at once."~\cite{Raaen.2015}. This statement seems to be no longer valid for current VR headsets like the \textit{Oculus Rift CV1} and \textit{HTC Vive}.

The measurement system is easy to set up and a feasible tool to evaluate distributed VR systems. It is open source, cheap and easy to build. It will help scientists and developers to better understand and improve distributed VR systems regarding local and remote latency.

In the future, we plan to enhance the measurement system further. The external hardware can, for example, be extended to determine local and remote haptic delays in VR systems, as well as binocular audio latencies as described by \textit{Stitt et al.}~\cite{Stitt.2016}. 

\VRARsetbibstyle
\bibliography{bibliography/bibliography}

\end{document}